\documentclass[twocolumn,showpacs,preprintnumbers,amsmath,amssymb]{revtex4}
\usepackage{graphicx}% Include figure files
\usepackage{dcolumn}% Align table columns on decimal point
\pdfoutput=1
\usepackage{bm}% bold math

\begin{document}

\preprint{}

\title{Cross-linking inhomogeneity in nano-composite hydrogels can be observed as sharp peaks by SAXS experiments under elongation.}% Force line breaks with \\

\author{Kengo Nishi$^1$}
\email{kengo.nishi@phys.uni-goettingen.de}
\author{Mitsuhiro Shibayama$^2$}
\email{sibayama@issp.u-tokyo.ac.jp}
\affiliation{ $^1$Third Institute of Physics-Biophysics, Georg August University, 37077 Goettingen, Germany, $^2$Institute for Solid State Physics, The University of Tokyo, 5-1-5 Kashiwanoha, Kashiwa, Chiba 277-8581, Japan}

\date{\today}% It is always \today, today,
             %  but any date may be explicitly specified

\begin{abstract}
We introduced silica nanoparticles into poly ($N$,$N$-dimethylacrylamide) gel, (PDAM-NC gel), and poly (acrylamide) gel (PAM-NC gel), and carried out SAXS measurements under uniaxial elongation. It is well known that PDAM chains are strongly adsorbed onto silica nanoparticles while PAM chains are not. Interestingly, we found from SAXS measurements that scattering profiles depend on the polymer-nanoparticle interaction. A four-spot pattern was observed in the 2D structure factors of PDAM-NC gel, which was assigned to a movement of the nanoparticles affinely under elongation. However, unexpectedly, we observed sharp peaks in the 2D structure factors of PAM-NC gel in the parallel direction to the stretching. These peaks appeared in much lower-$q$ region than the theoretical prediction of affine deformation of nanoparticles. We conjecture that these peaks do not correspond to the correlation of individual particles but to the correlation of high cross-linking regions because the interaction between PAM chains and silica nanoparticles is relatively weak and the displacements of silica nanoparticles are sensitive to spatial cross-linking inhomogeneity.   
\end{abstract}

% PACS, the Physics and Astronomy
                             % Classification Scheme.
%\keywords{Suggested keywords}%Use showkeys class option if keyword
                              %display desired
\maketitle

\section{\label{sec:level1} Introduction} 
Hydrogels are soft elastic solids and contain a large portion of water. By taking advantage of the high water absorption and elastic properties, hydrogels are applied to many kinds of industrial products such as diapers, drug reservoirs, and contact lenses. Though hydrogels have these characteristic properties which other materials do not have, industrial application is still limited due to their low mechanical strength. Because this low mechanical strength is thought to be related to cross-linking inhomogeneity, the relationship between cross-linking inhomogeneity and mechanical properties has been investigated by scattering experiments\cite{Mallam,Moussaidt,shibayama}, AFM\cite{Suzuki,Krzeminski,Nguyen}, microrheology\cite{Mason,Valentine,Penaloza}, and other techniques\cite{Kannurpatti,Guo}.

One of the most effective methods to enhance the mechanical properties of hydrogels is the introduction of fillers to polymeric materials such as carbon blacks\cite{Guth}, silica particles\cite{Petit}, carbon nanotubes\cite{CNT,Ajayan} and inorganic clays\cite{Haraguchi}. It is well known that the mechanical performance remarkably enhances by the introduction of fillers because of the interaction between fillers and polymers. Therefore, proper and better understanding of the interaction between fillers and polymers is necessary not only from basic science but also from industrial points of view.

In order to study the interaction between fillers and polymers, we focused attention on  nano-composite poly ($N$,$N$-dimethylacrylamide) gels (PDAM gel) which are developed by Lin et al\cite{Lin1,Lin2,Rose}. They showed an enhancement of mechanical properties by introducing silica nanoparticles into PDAM gels because of the strong adsorption of PDAM chains onto silica nanoparticles. Furthermore, they also reported that the introduction of silica nanoparticles into PAM gels does not lead to any significant reinforcement in the mechanical properties as opposed to PDAM gels because the interaction between silica nanoparticles and PAM chains is weaker than PDAM chains. Thus, in this study, we investigated the effect of the interaction of silica/polymer on the internal structure of nanocomposite hydrogels by SAXS. From SAXS experiments, we found that scattering profiles depend on the polymer-nanoparticle interaction. Furthermore, we unexpectedly observed a sharp peak in nano-composite PAM gel in very low-$q$ region, which may correspond to cross-linking inhomogeneity.

\section{\label{sec:level2} Experiment} 

\subsection{Materials} 
Acrylamide (AM, Wako), $N$,$N$-Dimethylacrylamide (DAM, Wako), potassium persulfate (KPS, Sigma Aldrich), $N$,$N$,$N$',$N$'-tetramethylethylenediamine (TEMED, Sigma-Aldrich), and $N$,$N$'-methylenebis-(acrylamide) (BIS, Sigma Aldrich) were used as received. The silica nanoparticles (Ludox TM-50 from Dupont) were obtained from Sigma Aldrich. From our preliminary SAXS experiment on dilute solution of silica nanoparticles, the mean radius of nanoparticles was estimated to be $\sim 13$ nm.

\begin{table*}[hbtp]
 \caption{Composition of Nano-Composite Hydrogels}
 \centering
\begin{tabular}{ c | c | c | c | c | c }
  Sample Name & silica weight fraction & $w_{water}$/g & $w_{silica}$/g & $w_{MBA}$/g & $w_{monomer}$/g\\
  \hline
  PAM-NC gel & 0.2 & 10 & 4.63 & 2.3 & 2.132 \\
  PDAM-NC gel & 0.2 & 10 & 4.33 & 2.3 & 2.974\\
\end{tabular}
\end{table*}

\subsection{Sample preparation} 
Compositions of the hydrogels are summarized in Table 1. The molar ratio of monomer to [TEMED] and to [KPS] was fixed at 100 to 1. The cross-linking density was also held constant as a cross-linker/monomer molar ratio of 3M/1.5mM. The amount of silica particles was fixed at 20 wt\%. A certain amount of KPS was dissolved in 2 ml deionized water. Separately, certain amounts of silica suspension, monomer, TEMED and BIS are dissolved in the certain amount of water. All solutions are deoxygenated by nitrogen bubbling for 15 min and degassed by vacuuming. This vacuum state was kept just until initiating free-radical polymerization. The polymerization was initiated by mixing two solutions under nitrogen conditions at room temperature and the mixture was transferred into fluorine-containing rubber molds which have dumbbell shape standardized as JIS K 6261-7 sizes (1mm thick).

\subsection{SAXS experiment} 
SAXS measurements were carried out  at the BL03XU beamline (Frontier Softmaterial Beamline (FSBL)) at SPring-8 that is located in Sayo, Hyogo, Japan. For PDAM-NC gel, a monochromated X-ray beam with a wavelength ($\lambda$) of 1.5 $\AA$ was used to irradiate the samples at room temperature, and the sample-to-detector distances was set to be 4 m. The scattered X-rays were counted by an imaging plate detector (R-AXIS VII++, Rigaku Corporation, Japan) with 3000 $\times$ 3000 pixel arrays and a pixel size of 0.1 mm pixel$^{-1}$. For PAM-NC gel, we used X-ray beam with a wavelength ($\lambda$) of 1.0 $\AA$ and 1.5 $\AA$ for the imaging plate detector and a CCD camera. In each measurement, samples were placed on a uniaxial stretching machine, where the sample strain were measured.

 In general, the SAXS intensity of particle systems is described by\cite{Guinier}
\begin{eqnarray}
I(\mbox{\boldmath $q$})=I_0P(\mbox{\boldmath $q$})S(\mbox{\boldmath $q$})
\end{eqnarray}
where $I_0$, $\mbox{\boldmath $q$}$, $P(\mbox{\boldmath $q$})$, and $S(\mbox{\boldmath $q$})$ are the zero-angle scattering intensity, the scattering vector, the form factor and the structure factor, respectively. Note that scattering intensities of nano-composite gels are more than 100 times higher than that of pure hydrogels without nanoparticles because the scattering length density of silica nanoparticle is much larger than that of polymers. (Data were not shown.) Thus, we assumed that scattering from polymers is negligibly small in the scattering profile of nano-composite gels and only silica particles contribute to the scattering profiles of nano-composite hydrogels. On the basis of this result, in order to obtain 2D structure factors of gels, we carried out SAXS experiments on a dilute solution of nanoparticles and the scattering profiles of the dilute solution of nano-composite gels are divided by that of nanoparticles by following the multiplication of a normalized constant to meet $S(\mbox{\boldmath $q$}) = 1$ in high-$q$ region.

 \section{\label{sec:level3} Results and Discussions}

 \begin{figure}[htbp]
\includegraphics[width=80mm]{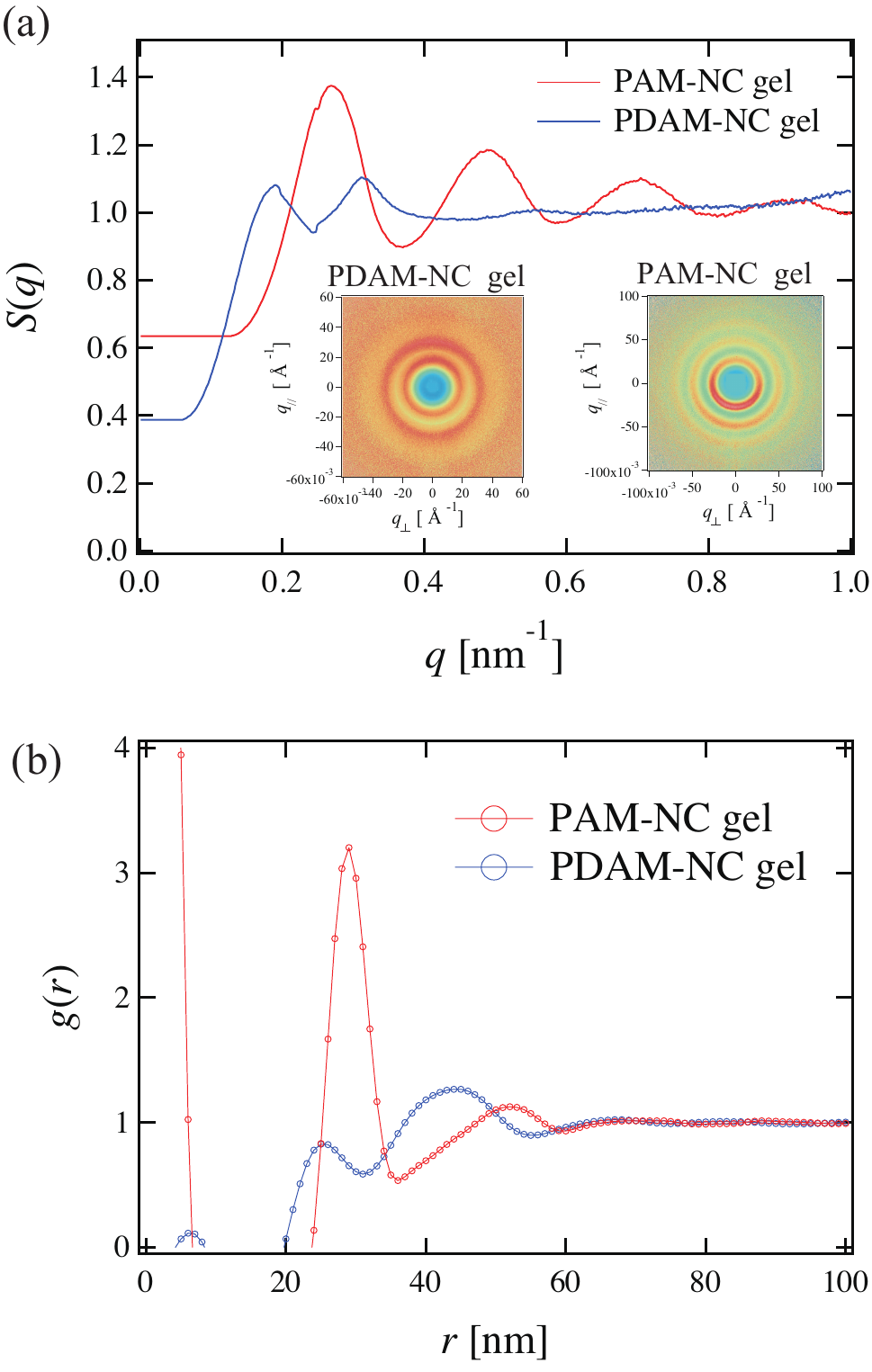}% Here is how to import EPS art
\caption{(a) 1D plot of structure factors for unstretched gels. (b) The pair distribution function for unstretched gels. The pair distribution function was evaluated from structure factor by Eq. (2). The 2D scattering intensity profiles are shown in the insets.}
\end{figure}
Let us start from SAXS data of PAM-NC gel and PDAM-NC gel in unstretched state. The structure factors for PAM-NC gel and PDAM-NC gel in unstretched state are shown in Fig. 1(a). The 2D scattering intensity profiles are shown in the insets. In order to evaluate the pair distribution function $g(r)$ , we cut the data in low-$q$ region and put the constant values based on extrapolation of structure factors. This treatment corresponds to removal of parasitic scattering or disregard of large aggregation which is out of our experimental range.  As shown in Fig. 1(a), the structure factor of PAM-NC gel is very different from that of PDAM-NC gel. We calculated the pair distribution function for gels from above structure factors in order to analyze the distribution of nanoparticles in gels by using the following equation\cite{silvia}.
 
\begin{eqnarray}
g(r)=1+\frac{1}{2\pi^2r{\rho_0}}\int^{\infty}_{0}{\hspace{0.5em}}dq(S(q)-1)\sin(qr)\exp\Bigl(-\frac{q^2}{2\sigma^2}\Bigr)\nonumber\\
\end{eqnarray} 
Here, $\rho_0$ is the number density of nanoparticles. The last exponential term is a window function to remove the termination ripples. $\sigma$ is the standard deviation and we set this value as $\sigma=q_{max}/2$ by following the previous works \cite{silvia}. The pair distribution functions of PAM-NC gel and PDAM-NC gel are shown in Fig. 1(b). We observed a large positive and negative values in the pair distribution function near $r=0$, which seems to be derived from the finite $q$-range in experimental data. In the case of PDAM-NC gel, we observed two peaks at 25 nm and 44 nm. When we take into account the fact that the radius of nanoparticle is  $\sim$13 nm, the peak at 25 nm indicates that some particles are directly-contacted. The peak at 44 nm may correspond to the distances between remaining nanoparticles which are homogeneously dispersed. As for PAM-NC gel, a strong peak can be observed at 30 nm. If we assumed that all nanoparticles are homogeneously dispersed in the gel, the distance between nearest neighbor particles is calculated to be 30.6 nm. Thus, nanoparticles in PAM-NC gel are homogeneously dispersed in the system, and the peak at 30 nm corresponds to the distance between nearest neighbor particles.
 
\begin{figure}[htbp]
\includegraphics[width=80mm]{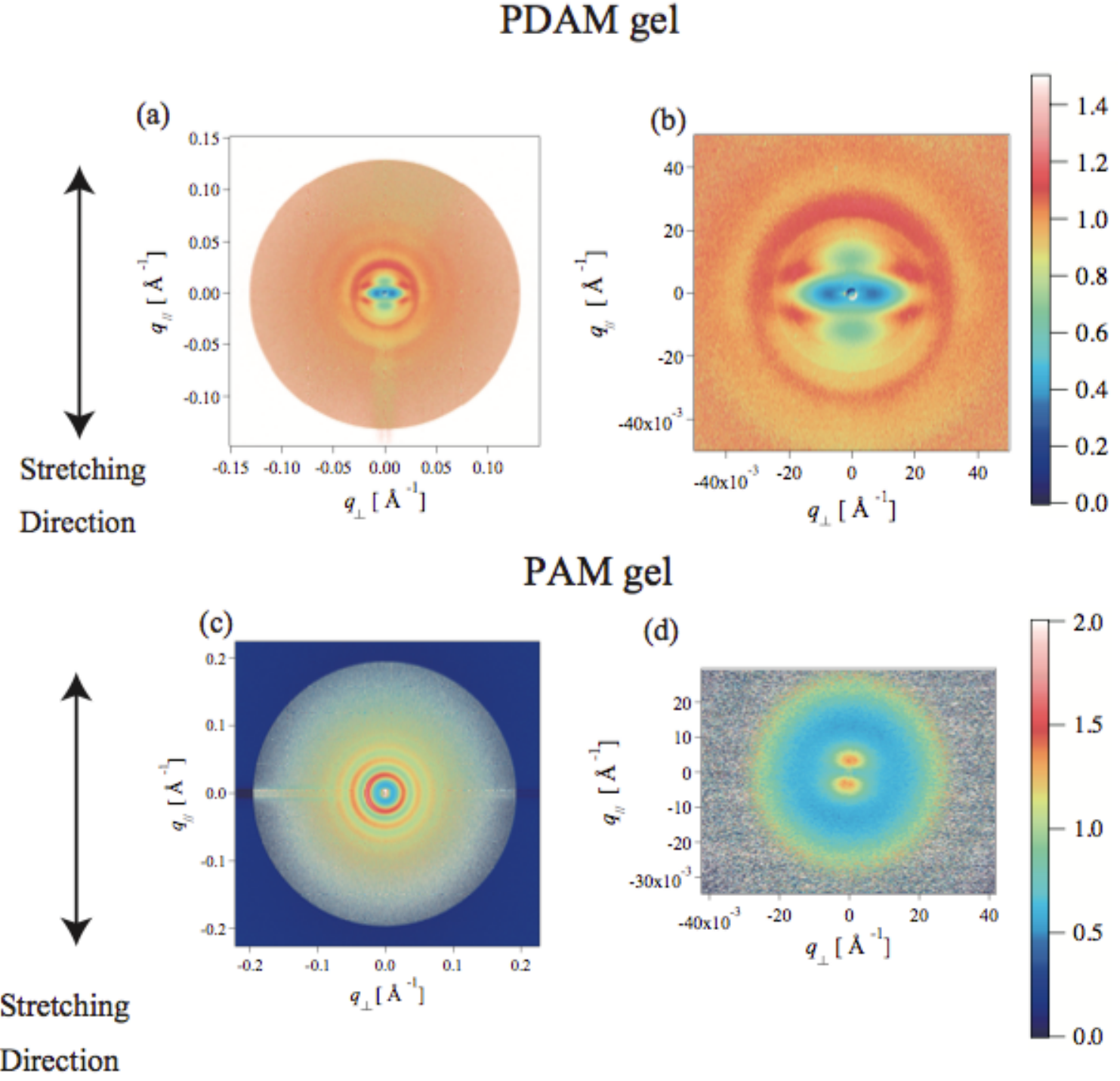}% Here is how to import EPS art
\caption{2D structure factors of gels. (a) 2D structure factors of PDAM-NC gel at $\lambda=2.03$. (b) A magnified figure of (a) in the vicinity of beam center. (c) 2D structure factors of PAM-NC gel at $\lambda=1.29$. This pattern was taken by using an imaging plate and the X-ray wavelength used was $1.0 \AA$. (d) 2D structure factors of PAM-NC gel at $\lambda=1.28$. This pattern was taken by using a CCD detector and the X-ray wavelength used was $1.5 \AA$.}
\end{figure}
We plotted 2D structure factors of PDAM gels for stretched states in Fig. 2(a). An anisotropic structure factor is observed when the gel is stretched.  In order to see the low-$q$ region in detail, we plotted a magnified figure of Fig. 2(a) in the vicinity of beam center in Fig. 2(b). As shown in Fig. 2(b), we observed a four spot pattern. Rose et al. conducted SANS experiments on a similar system (PDAM-NC gels) to our work in detail\cite{Rose}. However, they did not observe such a four-spot pattern but "abnormal butterfly pattern". This difference may be derived from the following points: (i) The wavelength distribution of the incident X-ray beam is narrower than that of neutron beam. Thus, SAXS has much higher resolution than SANS. (ii) The polymer concentration of PDAM-NC gel in the previous work\cite{Rose} is half of our experiment.  (iii) SANS detected not only silica particle but also polymers, while SAXS detected almost only silica particles. Though we cannot clearly deduce the reason for this difference, the difference of polymer concentration may be essential because four spot pattern is frequently observed in the nano-composite elastomers and rubbers\cite{Rharbi,Shinohara,Ikeda}. The origins of this four-spot pattern have already been discussed\cite{Rharbi}. They proposed two processes, i.e., some long aggregates buckle under the lateral compression, or the space in the perpendicular direction to the stretching between two aggregates is filled up by another aggregate that forced between them. However, at this stage, we will not discuss the origin of the four-spot pattern further and we will only show the analysis along the stretching direction. On the other hand, 2D structure factors of PAM-NC gel ($\lambda=1.29$) is completely different from those of PDAM-NC gel as shown in Fig. 2(c). Note that we use X-ray with a wavelength of 1.0 $\AA$ for this result. In the intermediate range, we observed a symmetric ring patterns. In the vicinity of low-$q$ limit, on the other hand, we observed a two-spot pattern. In order to see this two-spot pattern more clearly, we carried out SAXS measurement on stretched PAM-NC gel  ($\lambda=1.28$) by using X-ray with a wavelength of 1.5 $\AA$ (Fig. 2(d)). The clear two-spot pattern can be observed in the parallel direction of stretching. Unexpectedly, these peaks seem to appear in much lower $q$-region than the theoretical prediction of affine deformation of nanoparticles.

\begin{figure}[htbp]
\includegraphics[width=80mm]{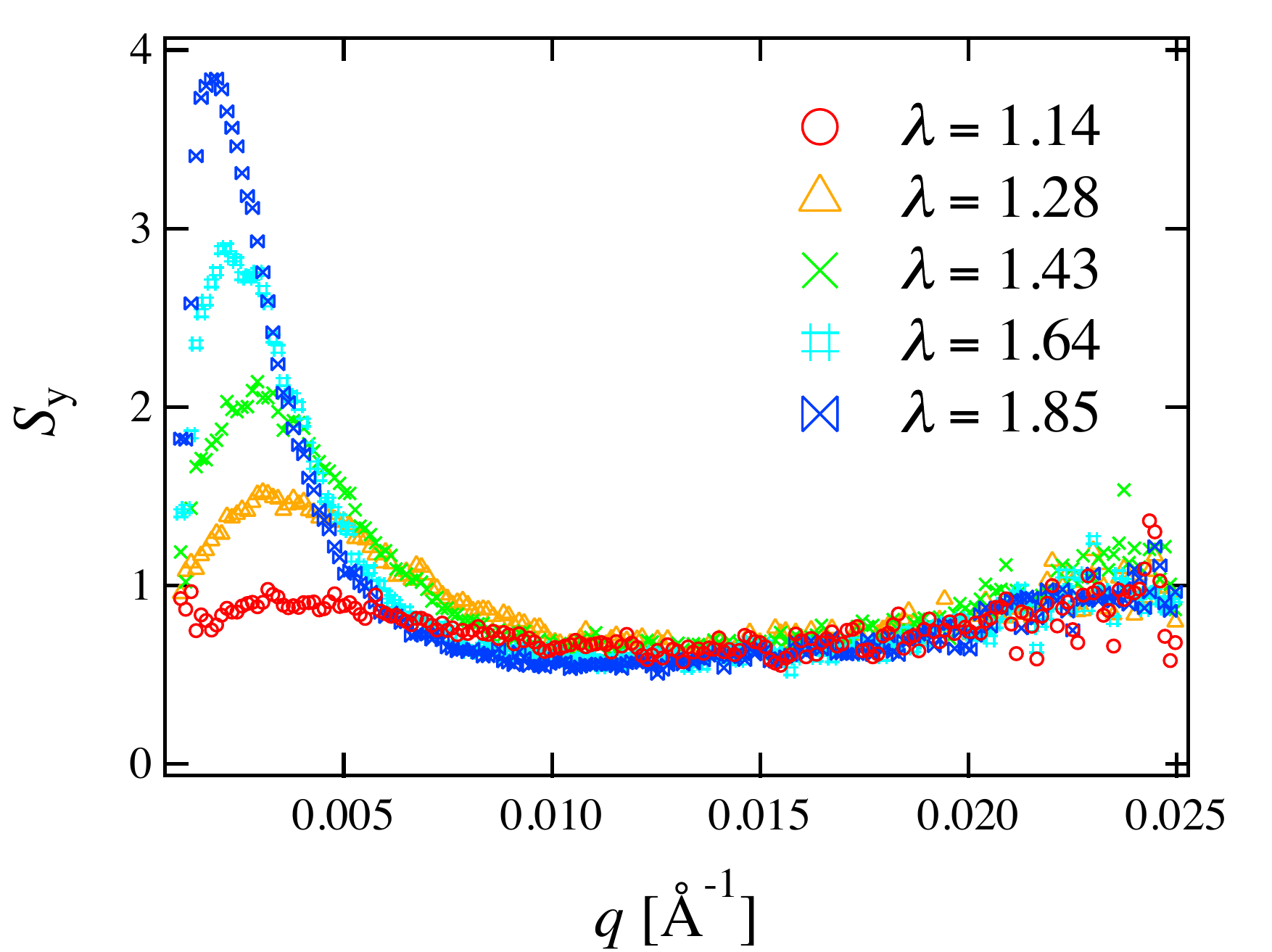}% Here is how to import EPS art
\caption{Structure factors of PAM-NC gel studied at different strtching ratios in the parallel direction of stretching.}
\end{figure}

 In order to investigate this peak in detail, we plotted slice images of 2D structure factors of PAM-NC gel at different stretching ratios in the parallel direction to the stretching in Fig. 3. As shown in Fig. 3, the peak becomes more intense and $q$-values of peak tops $q_{y,peak}$ decrease as the stretching ratio increases. This result suggests that distance between particles becomes systematically larger as  the stretching ratio increases.

\begin{figure}[htbp]
\includegraphics[width=80mm]{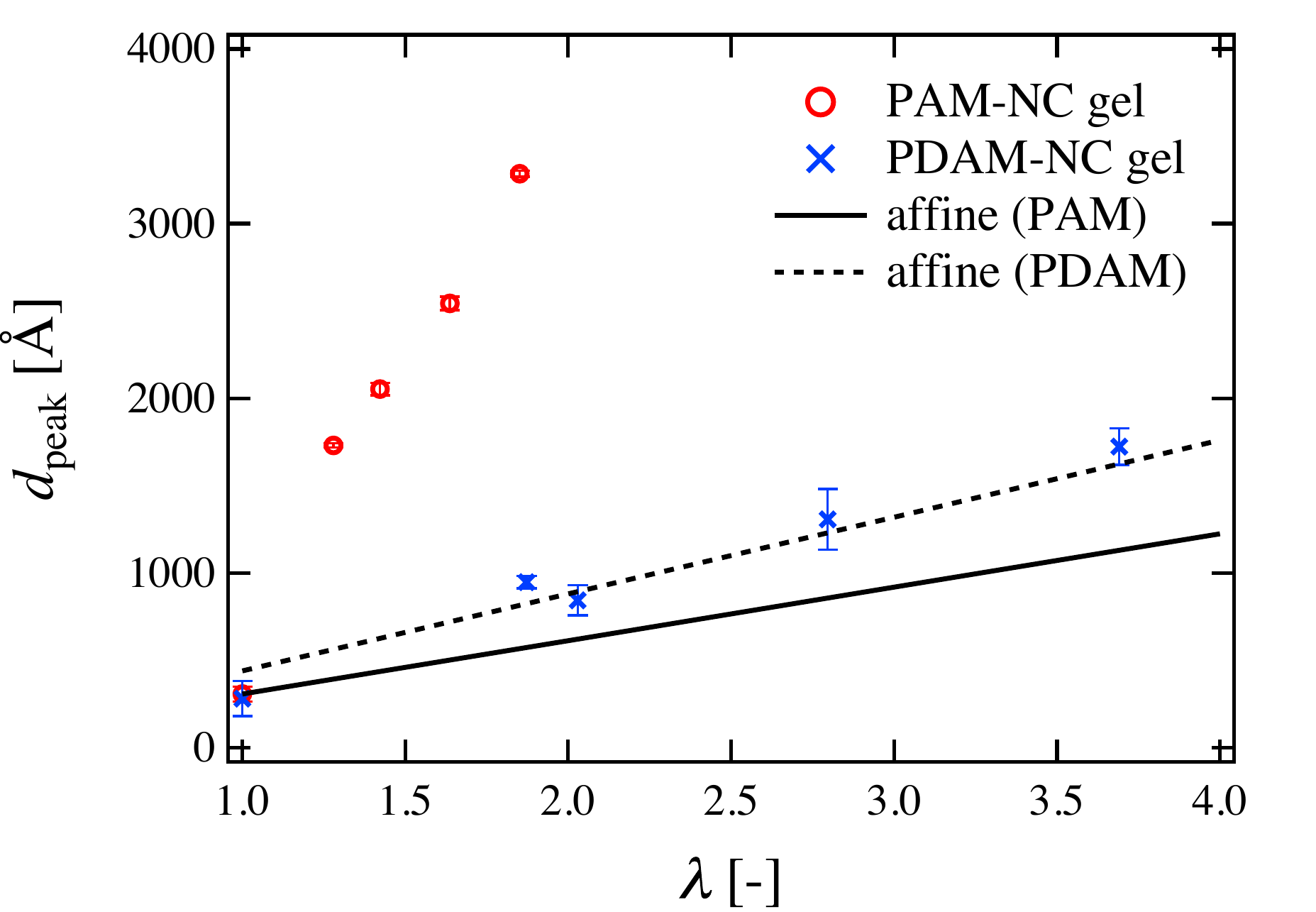}% Here is how to import EPS art
\caption{Stretching ratio dependence of the interparticle distances evaluated from $q_{y,peak}$ in the parallel direction of stretching. The solid line is a theoretical line calculated for affine deformation in the parallel direction. }
\end{figure}

We evaluated the distances between particles ($d_{peak}$) from the peak position ($q_{y,peak}$) in the low-$q$ region, such as $d_{peak}=2\pi/q_{y,peak}$. The solid and dotted lines shown in Fig. 4 are theoretical lines calculated for affine deformation in the parallel direction to the stretching for PAM-NC gel and PDAM-NC gel, respectively. As for the evaluation of $q_{y,peak}$ for PDAM-NC gel, we focused on the distance between nanoparticles which are not directly-attached each other and homogeneously dispersed in the system. Furthermore, we selected the center of each peak of the four-spot pattern and used its $y$-coordinate in the same way as previous works\cite{Shinohara,Ikeda}. As shown in Fig. 4, nanoparticles are moved in an affine way in the case of PDAM-NC gel. However, in the case of PAM-NC gel, $d_{peak}$ systematically increases but is remarkably larger than the prediction of affine deformation. This discrepancy cannot be understood if we assume that this peak corresponds to correlations between nearest neighbor particles.

\begin{figure}[htbp]
\includegraphics[width=80mm]{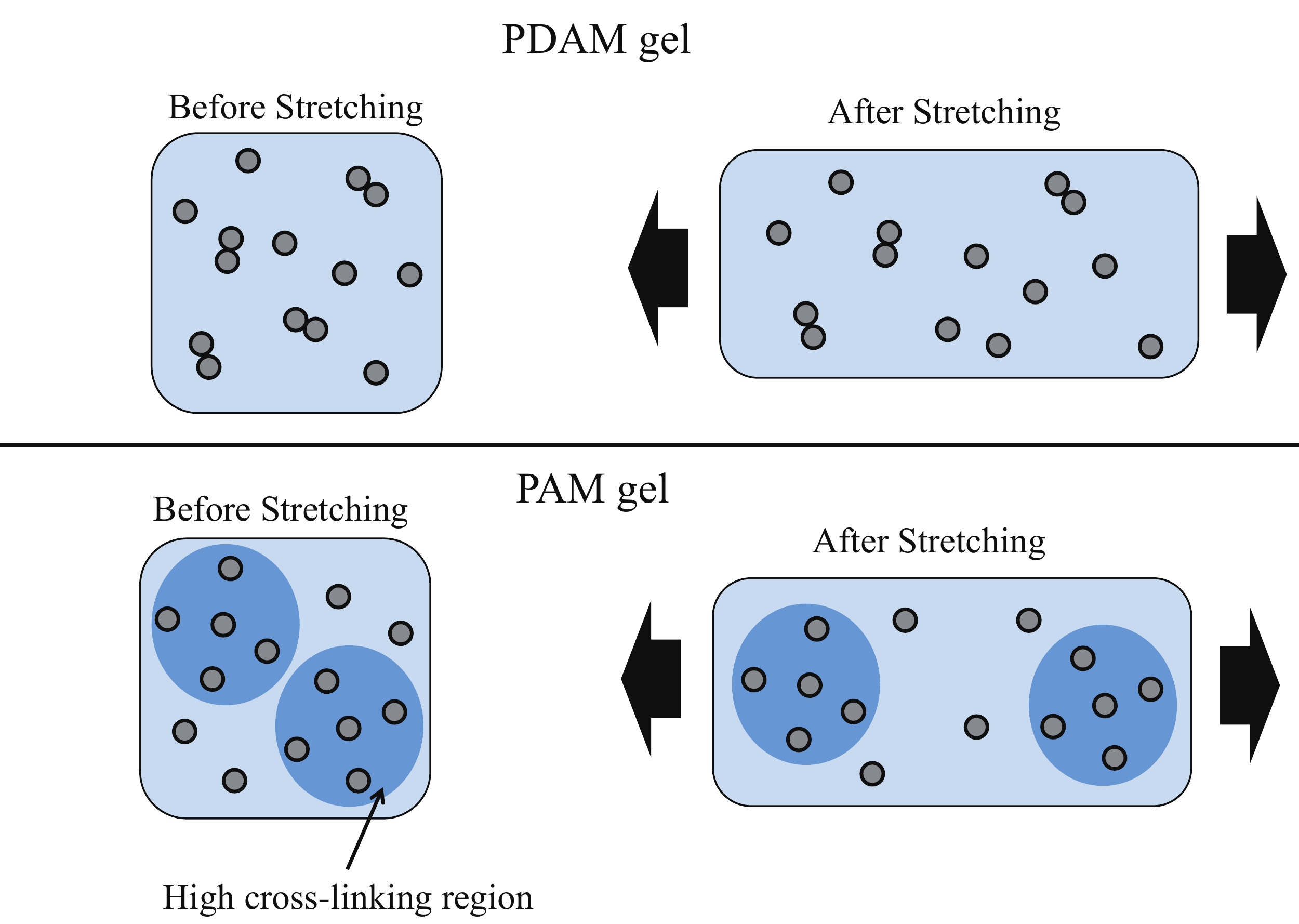}% Here is how to import EPS art
\caption{Schematic images for the stretching mechanism of PDAM-NC gel and PAM-NC gel. Small circles and large hatched circles  denote silica particles and high cross-linking regions, respectively. }
\end{figure}

In order to understand the origin of this peak, we depicted a schematic illustration for the deformation mechanism inside the PDAM-NC and PAM-NC gels. As shown in Fig. 5, some nanoparticles in PDAM-NC gels are directly-contacted  before stretching. When PDAM-NC gel is uniaxially-stretched, nanoparticles which are not directly-contacted each other move in an affine way. On the other hand, in the case of PAM-NC gel, all nanoparticles are homogeneously dispersed in the system in the unstretched state although there may inherently exist cross-linking heterogeneity in the gels. When PAM-NC gel is stretched, high cross-linking region is hardly deformed, while low cross-linking region is thought to be largely deformed. Thus, the difference of  nanoparticles density between high cross-linking region and low cross-linking region increases and high cross-linking region is exaggerated as increasing the stretching ratio as depicted in Fig. 5. According to the previous work \cite{Lin2}, an introduction of silica nanoparticles into PAM-NC gels does not lead to any significant effect in the mechanical properties, such as elastic modulus, hysteresis and fracture toughness, while an introduction of silica nanoparticles into PDAM-NC gels largely increases the mechanical properties of hydrogels. The key difference between the two hybrid gels seems to be that PAM chains are not adsorbed on silica while PDAM are. By taking into account these results, each nanoparticle can move in the affine way when polymer/filler interaction is strong and polymers chains are adsorbed onto the fillers such as PDAM-NC gel and ordinary rubber. On the other hand, the cross-linking inhomogeneity can be clearly reflected in the displacements of nanoparticles under stretching when polymer/filler interaction is weak and polymer chains are not adsorbed on the fillers such as PAM-NC gel. According to classical theories of rubber elasticity such as affine model\cite{Flory}, the magnitude of the deformation of all polymer is assumed to be derived from the macroscopic strain in an affine way. However, the above experimental results clearly indicate that the magnitude of the deformation of polymers seems to depend on the cross-linking inhomogeneity. 

In addition, this experimental finding may lead to an establishment of the evaluation method of cross-linking inhomogeneity in rubbers and gels. Stated another way, by introducing nanoparticles into polymeric materials which do not interact with the nanoparticles in the same way as microrheology\cite{Mason,Gittes,Schnurr} and conducting SAXS experiments on the materials under uniaxial elongation, we are able to systematically evaluate cross-linking inhomogeneity in the polymeric materials and investigate the correlation between cross-linking inhomogeneity and mechanical properties.

 \section{\label{sec:level4} Conclusion} 
 We conducted SAXS measurements on PDAM-NC gel and PAM-NC gel under uniaxial elongation and evaluated 2D structure factors. Interestingly, we found from SAXS measurements that scattering profiles depend on the polymer-nanoparticle interaction. A four-spot pattern was observed for PDAM-NC gel as in the case of nano-composite elastomers and rubbers, while sharp peaks were observed in the parallel direction to the stretching in the very low-$q$ region in PAM-NC gel. We evaluated stretching ratio dependence of interparticle distances from these scattering peaks, and found that nanoparticles in PDAM-NC gel are moved in an affine way. However, in the case of PAM-NC gel, the interparticle distances from these scattering peaks are much larger than the theoretical prediction of affine deformation of nanoparticles. This is because the scattering peak in PAM-NC gel does not correspond to the correlation peak of individual particles but to the correlation peak of high cross-linking region. In order to test this hypothesis, 2-dimensional fourier transformation and 2-dimensional reverse monte carlo simulation must be useful, which will be our future works.

 \section{\label{sec:level5} Author contributions} 
K.N. initiated and conceived the project, designed and conducted the experiments, interpreted results, and wrote the manuscript. M.S. gave a general discussion, supervised the project and also wrote the manuscript.

\section{\label{sec:level6} Acknowledgement} 
We would like to thank Atsushi Izumi of Advanced Technologies R$\&$D Laboratory, Sumitomo Bakelite Co., Ltd. for a kind support for SAXS experiments. We also would like to thank Xiang Li, Kazu Hirosawa, Yasuyuki Shudo of Neutron Science Laboratory, Institute for Solid State Physics, The University of Tokyo for a kind support for SAXS experiments. K.N. acknowledges the support from Research Fellowship for Young Scientists of the Japan Society for the Promotion of Science. The SAXS experiments were performed at the second hutch of SPring-8 BL03XU (Frontier Softmaterial Beamline (FSBL)) constructed by the Consortium of Advanced Softmaterial Beamline.

\vskip -\lastskip 

\vskip -10pt


\begin{thebibliography}{27}
\bibitem{Mallam}S. Mallam, F. Horkay, A. M. Hecht, E. Geissler, Macromolecules, 1989, 22, 3356-3361.
\bibitem{Moussaidt}A. Moussaidt, S. J. Candau, J. G. H. Joosten, Macromolecules, 1994, 27, 2102-2110.
\bibitem{shibayama} M. Shibayama, Macromol. Chem. Phys., 1998, 199, 1-30.
\bibitem{Suzuki}A. Suzuki, M. Yamazaki, Y. Kobiki, J. Chem. Phys., 1996, 104, 1751-1757.
\bibitem{Krzeminski}M. Krzeminski, M. Molinari, M. Troyon, X. Coqueret, Macromolecules, 2010, 43, 8121-8127.
\bibitem{Nguyen}H. K. Nguyen, M. Ito, S. Fujinami, K. Nakajima, Macromolecules, 2014, 47, 7971−7977.
\bibitem{Mason} T. G. Mason, D. A. Weitz, Phys. Rev. Lett., 1995, 74(7), 1250-1253.
\bibitem{Valentine}M. T. Valentine, P. D. Kaplan, D. Thota, J. C. Crocker, T. Gisler, R. K. Prud'homme, M. Beck, D. A. Weitz, Phys. Rev. E, 2001, 64, 061506.
\bibitem{Penaloza}David P. Penaloza, Jr., Atsuomi Shundo, Keigo Matsumoto, Masashi Ohno,Katsuaki Miyaji, Masahiro Gotoad, Keiji Tanaka, Soft Matter, 2013, 9, 5166.
\bibitem{Kannurpatti}A. R. Kannurpatti, K. J. Anderson, J. W. Anseth, C. N. Bowman, J. Polym. Sci. Pt. B Polym. Phys., 1997, 35, 2297-2307.
\bibitem{Guo}Z. Guo, H. Sautereau, D. E. Kranbuehl, Macromolecules, 2005, 38, 7992-7999.
\bibitem{Guth}E. Guth, J. Appl. Phys., 1945, 16(20), 20-25.
\bibitem{Petit} L. Petit, L. Bouteiller, A. Brulet, F. Lafuma, D. Hourdet, Langmuir, 2007, 78, 56005-56010.
\bibitem{CNT}L. S. Schadler, S. C. Giannaris, P. M. Ajayan, Appl. Phys. Lett., 1998, 73, 3842-3844.
\bibitem{Ajayan} P. M. Ajayan, L. S. Schadler, C. Giannaris, A. Rubio, Adv. Mater., 12(10), 750-753.
\bibitem{Haraguchi}K. Haraguchi, T. Takehisa, Adv. Mater., 2002, 14(16), 1120-1124.
\bibitem{Lin1}W. C. Lin, W. Fan, A. Marcellan, D. Hourdet, C. Creton, Macromolecules, 2010, 43, 2554-2563.
\bibitem{Lin2}W. C. Lin, A. Marcellan, D. Hourdet, C. Creton, Soft Matter, 2011, 7, 6578-6582.
\bibitem{Rose}S. Rose, A. Dizeux, T. Narita, D. Hourdet, C. Creton, Macromolecules,2013, 46, 4095-4104.
\bibitem{Guinier}A. Guinier, G. Fournet, Small-Angle Scattering of X-rays. Wiley, New York, 1955.
\bibitem{silvia}D. S. Silvia, Elementary Scattering Theory: For X-Ray and Neutron Users; Oxford University Press: NewYork, 2011.
\bibitem{Rharbi}Y. Rharbi, B. Cabane, A. Vacher, M. Joanicot, F. Boue, Europhys. Lett., 1999, 46(4), 472-478.
\bibitem{Shinohara}Y. Shinohara, H. Kishimoto, K. Inoue, Y. Suzuki, A. Takeuchi, K. Uesugi, N. Yagi, K. Muraoka, T. Mizoguchi, Y. Amemiya, J. Appl. Cryst., 2007, 40, s397-s401.
\bibitem{Ikeda}Y. Ikeda, Y. Yasuda, S. Yamamoto, Y. Morita, J. Appl. Cryst., 2007, 40, s549-s552. 
\bibitem{Flory}P. J. Flory, Principles of Polymer Chemistry; Cornell University Press: Ithaca, NY, and London, 1953.
\bibitem{Gittes}F. Gittes, B. Schnurr, P. D. Olmsted, F. C. MacKintosh, C. F. Schmidt, Phys. Rev. Lett., 1997, 79(17), 3286-3289.
\bibitem{Schnurr}B. Schnurr, F. Gittes, F. C. MacKintosh, C. F. Schmidt, Macromolecules, 1997, 30, 7781-7792.
\end{thebibliography}
\end{document}